\documentclass[prd,nofootinbib,a4paper,twocolumn,showkeys]{revtex4}

\usepackage{graphicx}
\usepackage{amsmath}
\usepackage{amssymb}
\usepackage{amsfonts}
\usepackage{latexsym}
\usepackage{mathrsfs}
\newcommand{\tr}{{\rm tr}}
\newcommand{\sign}{{\rm sign}}

\newcommand{\txtint}{{\textstyle\int}}

\begin{document}

\title{Deriving Boltzmann Equations from Kadanoff--Baym Equations\\ in Curved Space--Time}

\author{A. Hohenegger}
\email[]{andreas.hohenegger@mpi-hd.mpg.de}

\author{A. Kartavtsev}
\email[]{alexander.kartavtsev@mpi-hd.mpg.de}

\author{M. Lindner}
\email[]{manfred.lindner@mpi-hd.mpg.de}

\affiliation{Max--Planck Institute f\"ur Kernphysik, Saupfercheckweg 1, 69117 Heidelberg, Germany}

\begin{abstract}
To calculate the baryon asymmetry in the baryogenesis 
via leptogenesis scenario one usually uses Boltzmann 
equations with  transition amplitudes computed in vacuum. 
However, the hot and dense medium and, potentially, the
expansion of the universe can affect the collision terms 
and hence the generated asymmetry. In this paper we 
derive the Boltzmann equation in the curved space-time from 
 (first-principle) Kadanoff--Baym equations.
As one expects from general considerations, the derived 
equations are  covariant generalizations of the 
corresponding equations in Minkowski space-time. 
We find that, after the necessary approximations have been performed, only the left-hand 
side of the Boltzmann equation depends on the space-time metric.
The amplitudes in the collision term 
on the right--hand side are independent of the metric, which 
justifies earlier calculations where this has been assumed implicitly. At tree level, 
the matrix elements coincide with those computed in vacuum. 
However, the loop contributions involve additional  integrals 
over the the distribution function. 
\end{abstract}

\maketitle

\section{Introduction}
As has been shown by A. Sakharov \cite{Sakharov:1967dj}, 
the observed baryon asymmetry of the universe can be generated 
dynamically, provided that the following three conditions are 
fulfilled: violation of baryon (or baryon minus lepton) number;
violation of \textit{C} and \textit{CP}; and deviation from 
thermal equilibrium. 

The third Sakharov condition raises  the question of how to 
describe a quantum system out of  thermal equilibrium. 
The usual choice is the Boltzmann equation \cite{Bernstein:1988,Groot:1980,
Cercignani:2002,Liboff:2003}. However, it is known to have several 
shortcomings. In particular classical Boltzmann equations 
neglect off--shell effects, introduce irreversibility and 
feature spurious constants of motion. 
A quantum mechanical generalization of the Boltzmann equation,
free of the mentioned problems, has been developed by 
L. Kadanoff and G. Baym \cite{KB:1962}. Direct numerical computations
demonstrate that already for simple systems far from thermal equilibrium 
the Kadanoff--Baym and Boltzmann equations do lead to 
quantitatively, and in some cases even qualitatively, different 
results \cite{Berges:2001fi,Aarts:2001qa,Lindner:2005kv,Lindner:2007am,
Berges:2005ai,Juchem:2004cs}. 
Studying processes responsible for the generation of the 
asymmetry in the framework of the Kadanoff--Baym formalism is 
therefore of considerable scientific interest.

The application of the Kadanoff--Baym equations to the computation of the 
lepton and baryon asymmetries in the leptogenesis scenario 
\cite{Fukugita:1986hr}  has been studied at different levels 
of approximation  by several authors \cite{Buchmuller:2000nd,DeSimone:2007rw} 
and lead to qualitatively new and interesting results.
However  issues related to the rapid expansion of the universe, 
which drives the required deviation from thermal 
equilibrium, have not been addressed there.
The modification of the Kadanoff--Baym formalism in  curved space--time 
has been considered in 
\cite{Calzetta:1986ey,Calzetta:1987bw,Ramsey:1997qc,Tranberg:2008ae},
where it was applied to a model with quartic self--interactions and a
 $O(N)$ model, though the dynamics of quantum field theoretical models with 
\textit{CP} violation  remained uninvestigated. 

Our goal is to develop a consistent  description of leptogenesis 
in the Kadanoff--Baym and Boltzmann approaches and to test approximations 
commonly made in the computation of the lepton and baryon asymmetries. 
In particular, we want to find out how the dense background plasma 
and the curvature of spacetime affect the collision terms of  
processes contributing 
to the generation and washout of the asymmetry, check the applicability 
of the real intermediate state subtraction procedure in the case 
of resonant leptogenesis \cite{Pilaftsis:2003gt,Pilaftsis:2005rv}, 
and  investigate the time dependence of the \textit{CP}--violating 
parameter in the expanding universe \cite{DeSimone:2007rw}. 

Since this is a rather ambitious goal, we 
first study a simple toy model of leptogenesis containing two 
real and one complex scalar fields, which mimic the heavy right--handed 
Majorana neutrinos and leptons respectively \cite{Hohenegger:2009a}. 
The peculiarities of the calculation, related to the presence of 
a gravitational field, are determined only by transformation 
properties of the quantum fields -- scalar fields in this case. For 
this reason, in the present paper, we use a model of a single real 
scalar field with quartic self--interactions, minimally
coupled to gravity, to illustrate the 
main points. That is, we use the Lagrangian
\begin{equation}
\label{lagrangian}
{\cal L}=
\frac12\partial^\mu \varphi\partial_\mu \varphi-\frac12 M^2 \varphi^2  
-\frac{\lambda}{4!}\varphi^4\,,
\end{equation}
which does also have the advantage, that one can compare the derived equations 
with their Minkowski space--time counterparts \cite{Lindner:2005kv} and 
with the results obtained in 
\cite{Calzetta:1986ey,Calzetta:1987bw,Ramsey:1997qc,Tranberg:2008ae}.
The formalism presented here will be used to analyze the toy model 
of leptogenesis.

The starting point of our analysis, which is manifestly covariant 
in every step, is the generating functional 
for the (connected) Green's functions. Performing a Legendre 
transformation we get the effective action, which we use to derive 
the Schwinger--Dyson equations in Sec.\,\ref{schwdyson}. These are equivalent to a system of  Kadanoff--Baym equations for the 
spectral function and the statistical propagator, which we derive 
in Sec.\,\ref{KB}. Employing a first--order gradient expansion and a 
Wigner transformation we are lead to a system of quantum kinetic equations 
which we study in Sec.\,\ref{QK}. Finally, neglecting the Poisson 
brackets and making use of the quasiparticle approximation, we obtain 
the Boltzmann equation in  Sec.\,\ref{BE}. 
\begin{itemize}
\item The Kadanoff--Baym equations and the derived Boltzmann equation are covariant generalizations of their Minkowski--space counterparts.

\item The space-time metric enters its left-hand side in the 
form of the covariant derivative, whereas the collision terms on the 
right-hand side are independent of the metric. 

\item At tree-level the collision terms coincide with those calculated 
in vacuum, whereas the loop corrections contain integrals over the 
distribution function. 

\item In the loop contributions one can clearly 
distinguish the initial, final and on--shell intermediate states,
which is not the case in the canonical formalism.
\end{itemize}
We discuss these results in more details and draw the conclusions 
in Sec.\,\ref{summary}.

\section{\label{schwdyson}Schwinger--Dyson equations}

In the derivation of the Schwinger-Dyson equations we employ results from \cite{Ramsey:1997qc,Basler:1991st,PhysRevD.35.3796}.
Our starting point is the generating functional for Green's 
functions with local and bi--local external scalar sources $J(x)$ 
and $K(x,y)$,
\begin{align}
\label{genfunct}
{\cal Z}[J,K]=
\txtint\mathscr{D}\varphi~\exp[i(S+J\varphi+{\textstyle\frac12}\varphi K \varphi)],
\end{align}
where the action $S$ is given by the integral of the Lagrange density 
over space. The Minkowski space--time volume element $d^4x$ is replaced in 
 curved space--time by the invariant volume element $\sqrt{-g}d^4x$, where 
$\sqrt{-g}$ is the square root of the determinant of the metric:
\begin{align*}
S&=\txtint \sqrt{-g} d^4x ~ {\cal L} \,.
\end{align*}
In the Friedmann--Robertson--Walker (FRW) universe we have $\sqrt{-g}=a^4(\eta)$, 
where $a$ is the  scale factor and $\eta$ de\-no\-tes conformal time.
The invariant volume element enters also in the scalar products 
of the  sources and the   field
\begin{subequations}
\begin{align}
J\varphi&\equiv\txtint\sqrt{-g}d^4 x J(x)\varphi(x)\,,
\\
\varphi K \varphi &\equiv {\textstyle\iint}\sqrt{-g}d^4 x \sqrt{-g}d^4 y 
~\varphi(x)K(x,y) \varphi(y)\,.
\end{align}
\end{subequations}
The functional integral measure is modified in  curved space--time as 
well. For scalar densities of zero weight it reads \cite{Basler:1991st}
\begin{align*}
\mathscr{D}\varphi=\prod_x d[(-g)^\frac14\varphi(x)]\,.
\end{align*}

\begin{figure}[h!]
\includegraphics[width=0.45\textwidth]{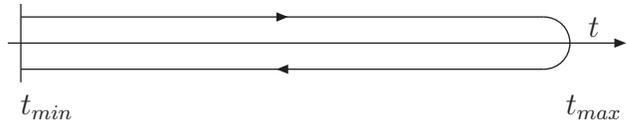}
\caption{\label{CTP}Closed real--time path $\cal C$. }
\end{figure}

The evolution of the quantum system out of thermal equilibrium 
is performed in the Schwinger--Keldysh formalism \cite{Schwinger:1960qe,
Keldysh:1964ud}. 
In this approach the field and the external sources 
are defined on the positive and negative branches of a closed 
real--time contour, see Fig.\,\ref{CTP}, the functions\footnote{In 
particular there are
two local  ($J_{+}$ and $J_{-}$) and four bi--local ($K_{++}$, 
$K_{+-}$, $K_{-+}$ and $K_{--}$) sources. Analogously, the field 
value on the two branches is denoted by $\varphi_{+}$ and 
$\varphi_{-}$ respectively, whereas the two--point function
components are denoted by $G_{++}$, $G_{+-}$, $G_{-+}$ and $G_{--}$ 
\cite{Chou:1984es}.} on the 
positive branch being independent\footnote{With the exception of 
the point $t=t_{max}$.} of the functions on the negative branch.
This applies also to the metric tensor, i.e.
$g^+_{\mu\nu}\neq g^-_{\mu\nu}$  in general.

In realistic models of leptogenesis the contribution of the heavy 
right--handed neutrinos to the energy density of the universe
is less than 5\% and can safely be neglected. In other words, 
leptogenesis takes place in a space--time with a metric, whose time 
development is  (in this approximation) independent of the decays of 
the right--handed neutrinos  and determined by the contributions of 
the ultrarelativistic standard model species. Correspondingly, in our analysis of the 
toy model of leptogenesis, we will also neglect the impact of the 
scalar fields on the expansion of the universe\footnote{
A theoretical analysis of the  back--reaction of the fields on the gravitational 
field has been performed in \cite{Calzetta:1986ey}. An  
analysis, with very interesting numerical results, of a model with quartic self--interactions in the 
Friedmann--Robertson--Walker universe has been carried out in 
\cite{Tranberg:2008ae}.}.
This implies in particular that the metric tensor on the positive
and negative branches is determined only by the external processes, 
and one can set $g^+_{\mu\nu}= g^-_{\mu\nu}=g_{\mu\nu}$. To 
shorten the notation we will also suppress the branch indices
of the scalar field and the  sources.

The existence of the two branches also affects the definition of the 
$\delta$ function: $\delta(x,y)$ is always zero if its arguments lie 
on different branches \cite{Danielewicz:1982kk}. In  curved space--time
it is further generalized  to fulfill the relation
\begin{equation}
{\txtint} d^4y \sqrt{-g}~ f(y)~ \delta^g(x,y)=f(x)\,,
\end{equation}
where the integration is performed over the closed contour.
The solution to this equation is given by \cite{Basler:1991st}
\begin{equation}
\label{generalizeddelta}
\delta^g(x,y)=(-g_x)^{-\frac14}\delta(x,y)(-g_y)^{-\frac14}\,.
\end{equation}
The generalized $\delta$ function is used to define functional 
differentiation in  curved space--time \cite{Zaidi:1983wf}
\begin{align}
\label{funcdiff}
\frac{\delta {\cal F}[\phi]}{\delta \phi(y)}\equiv 
\lim\limits_{\varepsilon\rightarrow 0}
\frac{{\cal F}[\phi(x)+\varepsilon\, \delta^g(x,y)]-{\cal F}[\phi(x)]}{\varepsilon}\,.
\end{align}
From the definition \eqref{funcdiff} it follows immediately that 
\begin{align}
\frac{\delta J(x)}{\delta J(y)} =\delta^g(x,y)~,\quad 
\frac{\delta K(x,y)}{\delta K(u,v)}  =\delta^g(x,u)\delta^g(y,v)\,.	
\end{align}

The functional derivatives of the generating functional for connected 
Green's functions
\begin{align}
\label{Wfunct}
{\cal W}[J,K]=-i\ln {\cal Z}[J,K]\,
\end{align}
with respect to the external sources read
\begin{subequations}
\begin{align}
\frac{\partial {\cal W}[J,K]}{\partial J(x)}&=\Phi(x)\,,\\
\frac{\partial {\cal W}[J,K]}{\partial K(x,y)}&=
{\textstyle\frac12}[G(y,x)+\Phi(x)\Phi(y)]\,,
\end{align}
\end{subequations}
where $\Phi$ denotes expectation value of the field and 
$G$ is the propagator. The effective action is the Legendre transform of the generating
functional for connected Green's functions,
\begin{align}
\Gamma[\Phi,G]\equiv {\cal W}[J,K]-J\Phi-{\textstyle\frac12}\tr[KG]
-{\textstyle\frac12}\Phi K \Phi\,.
\end{align}
Its functional derivatives with respect to the expectation value 
and the propagator reproduce the external sources:
\begin{subequations}
\label{DiffGamma}
\begin{align}
\frac{\delta \Gamma[G,\Phi]}{\delta \Phi(x)}&=-J(x)
-\txtint\sqrt{-g}d^4z~ K(x,z)\Phi(z)\,,\\
\frac{\delta \Gamma[G,\Phi]}{\delta G(x,y)}&=
-{\textstyle\frac12}K(y,x)\,.
\end{align}
\end{subequations}
Next, we shift the field by its expectation value 
\[
\varphi\rightarrow\varphi+\Phi\,.
\]
The action can then be written as a sum of two terms
\begin{align}
S[\varphi]\rightarrow S_{cl}[\Phi]+S[\varphi,\Phi]\,.
\end{align}
$S_{cl}$ denotes the classical action, which depends only on 
$\Phi$, whereas $S[\varphi,\Phi]=S_0[\varphi]+
S_{int}[\varphi,\Phi]$ contains terms quadratic, cubic 
and quartic in the shifted field $\varphi$. The free field action 
can be written in the form
\begin{align}
S_0=
{\textstyle\frac12\iint} \sqrt{-g}_x d^4x  \sqrt{-g}_y d^4y 
\, \varphi  \left(i\mathscr{G}^{-1}\right)\varphi\,,
\end{align}
where $\mathscr{G}^{-1}$ is the zero--order inverse propagator
\begin{align}
\hspace{-1mm}
\mathscr{G}^{-1}(x,y)=i(\square_x+M^2)\,\delta^g(x,y),\,
\square_x\equiv g_{\mu\nu}\nabla^\mu_x\nabla^\nu_x\,.
\end{align}
Since the integration measure in the path integral is translationally
invariant, the effective action can be rewritten in the form
\begin{align}
\Gamma[\Phi,G]=&-i\ln \txtint \mathscr{D}\varphi\exp [i(S %[\varphi,\Phi]
+
J\varphi+{\textstyle\frac12}\varphi K\varphi)]
\nonumber\\
&
+S_{cl}[\Phi]-{\textstyle\frac12}\tr[KG]\,.
\end{align}
Now we tentatively write the effective action in the form
\begin{align}
\label{G2deff}
\Gamma[\Phi,G]&\equiv S_{cl}[\Phi]+{\textstyle\frac{i}{2}}
\ln\det\left[G^{-1}\right]+{\textstyle\frac{i}2}\tr\left[\mathscr{G}^{-1}G\right]\nonumber\\
&\hspace{0.4cm}+\Gamma_2[\Phi,G]\,,
\end{align}
defining the functional $\Gamma_2$. The third term on the 
right--hand side is defined by 
\begin{align*}
\tr\left[\mathscr{G}^{-1} G\right]\equiv {\textstyle\iint}
\sqrt{-g_x}d^4 x \sqrt{-g_y}d^4 y 
~\mathscr{G}^{-1}(x,y) G(y,x)\,,
\end{align*}
whereas the second term on the right--hand side is defined by 
the path integral 
\begin{align*}
\det\left[\frac{G^{-1}}{2\pi}\right]\equiv\txtint \mathscr{D}\varphi
\exp\left(\varphi\, {G^{-1}} \varphi\right)\,.
\end{align*}
Using \eqref{DiffGamma} we can find the functional derivatives 
of $\Gamma$. 
Differentiation of $\tr\left[\mathscr{G}^{-1} G\right]$ with respect to 
$G$ is straightforward and gives
\begin{align}
\frac{\delta}{\delta G(x,y)}\tr\left[\mathscr{G}^{-1} G\right]=\mathscr{G}^{-1}(y,x)\,.
\end{align}
To calculate the functional derivative of $\ln\det\left[G^{-1}\right]$
we  take into account that in  curved space--time
\begin{align}
\label{inverseprop}
\txtint\sqrt{-g}\,d^4z\, G^{-1}(u,z)G(z,v)=\delta^g(u,v)~.
\end{align}
After some algebra and use of \eqref{inverseprop} we obtain a 
result analogous to that in  Minkowski space--time 
\begin{align}
\frac{\delta}{\delta G(x,y)}\ln\det \left[G^{-1}\right]=-G^{-1}(y,x)\,.
\end{align}
The functional derivative of \eqref{G2deff} with respect to $G$ then reads
\begin{align}
\label{G2diff}
\frac{\delta \Gamma[G,\Phi]}{\delta G(x,y)}=&
-{\textstyle\frac{i}{2}}G^{-1}(y,x)+{\textstyle\frac{i}2}\mathscr{G}^{-1}(y,x)
+\frac{\delta \Gamma_2[G,\Phi]}{\delta G(x,y)}\nonumber\\
=&-{\textstyle\frac12}K(y,x)
\end{align}
Solving \eqref{G2diff} with respect to $K$ and substituting 
it into \eqref{G2deff} we can rewrite the effective action in the form 
\begin{align}
\Gamma_2[G&,\Phi]=-i\ln\int \mathscr{D}\varphi \exp\left[
i\left(S+J\varphi-\varphi \frac{\delta \Gamma_2}{\delta G} \varphi\right)
\right]\nonumber\\
&+\tr\left[\frac{\delta \Gamma_2}{\delta G} G\right]
-{\textstyle\frac{i}{2}}\ln\det \left[G^{-1}\right]+const.\,,
\end{align}
where again $S=S_0+S_{int}$, but now with $S_0$ given by
\begin{align}
S_0={\textstyle\frac12\iint} \sqrt{-g}_x d^4x  \sqrt{-g}_y d^4y 
\, \varphi  \left(iG^{-1}\right)\varphi\,.
\end{align}
This implies that $i\Gamma_2$ is the sum of all 2PI vacuum diagrams 
with vertices as given by ${\cal L}_{int}$ and internal lines 
representing the complete connected propagators $G$ \cite{PhysRevD.10.2428}.

Physical situations correspond to vanishing sources. Introducing 
the self--energy 
\begin{align}
\label{selfenergydeff}
\Pi(x,y)\equiv 2i\frac{\delta \Gamma_2[G,\Phi]}{\delta G(y,x)}\,,
\end{align}
we can then rewrite \eqref{G2diff} in the form
\begin{align}
\label{SDeqs}
G^{-1}(x,y)=\mathscr{G}^{-1}(x,y)-\Pi(x,y)\,.
\end{align}
Thus the above calculation yields the Schwinger--Dyson (SD) equation. 
Let us note that the derived equation has exactly the same form as in 
Minkowski space--time.

\section{\label{2PI}2PI effective action}
\begin{figure}[!ht]
\begin{center}
\includegraphics[width=0.45\textwidth]{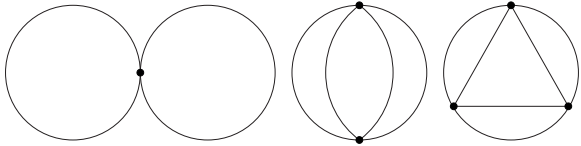}
\end{center}
\caption{\label{diagrams}Two--, three--, and four--loop contributions to the 
2PI effective action.}
\end{figure}

The structure of the Schwinger--Dyson equation is 
determined only by the particle content of the model 
(here a single real scalar field) and completely 
independent of the particular form of the interaction 
Lagrangian. The latter determines the form of the 2PI effective action.
The lowest order contribution is due to the two--loop 
diagram in Fig.\,\ref{diagrams},
which only takes into account local effects and cannot describe 
thermalization. Thus one usually also considers the three--loop 
diagram, which describes $2\leftrightarrow 2$ scattering. 
In addition we take into account the four--loop contribution. 
As is demonstrated below, in the Boltzmann approximation it describes 
the  one--loop correction for $2\leftrightarrow 2$ scattering.
The resulting expression for the effective action is 
similar to that given in \cite{Aarts:2001qa,Lindner:2005kv,Ramsey:1997qc,Berges:2004yj}:
\begin{align}
i\Gamma_2[G]&=\sum\limits_n i\Gamma^{(n)}_2[G],\\
i\Gamma^{(2)}_2[G]&=-\,\frac{i\lambda}{8}\txtint \sqrt{-g}_xd^4x\, G^2(x,x),\nonumber\\
i\Gamma^{(3)}_2[G]&=-\,\frac{\lambda^2}{48}\txtint \sqrt{-g}_xd^4x
\sqrt{-g}_yd^4y\, G^2(x,y)G^2(y,x),\nonumber\\
i\Gamma^{(4)}_2[G]&=\,\frac{i\lambda^3}{48}\txtint \sqrt{-g}_xd^4x
\sqrt{-g}_yd^4y\sqrt{-g}_zd^4z\nonumber\\
&\hspace{30mm}
\times G^2(y,x)G^2(x,z)G^2(z,y).\nonumber
\end{align}
Note, however, the presence of the $\sqrt{-g}$ factors which ensure 
invariance of the effective action under coordinate transformations. 

Using the definition of the self--energy \eqref{selfenergydeff} 
and the functional differentiation rule in  curved space---time 
we obtain 
\begin{align}
\label{selfenergy}
\Pi(x,y)=&\sum\limits_n \Pi^{(n)}(x,y)\,,\\
\Pi^{(2)}(x,y)=& -i\delta^g(x,y)\frac{\lambda}{2}G(x,x)\,,\nonumber\\
\Pi^{(3)}(x,y)=&-\frac{\lambda^2}{6}G(y,x)G(x,y)G(x,y)\,,\nonumber\\
\Pi^{(4)}(x,y)=&\frac{i\lambda^3}{4}G(y,x)\txtint\sqrt{-g}_zd^4z
G^2(x,z)G^2(z,y)\,.\nonumber
\end{align}
It is worth mentioning that the appearance of the generalized $\delta$ function 
in the first {\em local} term is a consequence 
of the form of the  effective action and the functional differentiation
rule \eqref{funcdiff}.  For  each vertex in the loop diagrams there is a 
corresponding integral in the effective action. Because of the 
appearance of the generalized $\delta$ functions two of the integrals 
can be carried out trivially after functional differentiation.
Further  integrals  persist in the self--energy. That is, four-- 
and higher--loop contributions to $\Pi(x,y)$  contain 
integrations over space--time  with the corresponding number of $\sqrt{-g}$ 
factors to ensure the invariance of the self--energy.

\section{\label{KB}Kadanoff--Baym equations}
Convolving the Schwinger--Dyson equations \eqref{SDeqs} with $G$ from
the right and using \eqref{inverseprop} we obtain
\begin{align}
\label{convolution}
i[\square_x+M^2]G(x,y)=&\delta^g(x,y)\nonumber\\
&+\txtint\sqrt{-g}d^4z\,\Pi(x,z)G(z,y)\,.	
\end{align}
Next, we define the spectral function 
\begin{align}
\label{Grho}
G_\rho(x,y)=i\langle [\varphi(x),\varphi(y)]_{-}\rangle\,,
\end{align}
and the statistical propagator 
\begin{align}
\label{Gf}
G_F(x,y)={\textstyle\frac12}\langle [\varphi(x),\varphi(y)]_{+}\rangle\,.
\end{align}
As is clear from the definitions, the statistical propagator 
of real scalar field is symmetric whereas the spectral function 
is antisymmetric with respect to permutation of its arguments.
For a real scalar field $G_F(x,y)$ and $G_\rho(x,y)$ are real--valued
functions \cite{Berges:2001fi}. The full Feynman propagator can 
be decomposed into a statistical and a spectral part
\begin{align}
\label{GfGr}
G(x,y)&=G_F(x,y)-{\textstyle\frac{i}{2}}\sign(x^0-y^0)G_\rho(x,y)\,.
\end{align}
Upon use of the $\sign$-- and $\delta$--function differentiation 
rules, the action of the $\square_x$ operator on the second term 
on the right--hand side of \eqref{GfGr} gives a product of $g^{00}\delta(x^0,y^0)$ and 
$\nabla^x_0G_\rho(x,y)$. Using the definition \eqref{Grho} and 
the canonical commutation relations in  curved space--time \cite{Isham:1978}
\begin{align}
\lim\limits_{y^0\rightarrow x^0}[\varphi(x^0,\vec{x}\,),\pi(x^0,\vec{y}\,)]_{-}&=i\delta (\vec{x},\vec{y}\,),
\end{align}
where\footnote{To simplify the calculation we set $g_{0i}=0$. The off--diagonal components 
of the metric tensor can always be set to zero by an appropriate choice of the coordinate system 
\cite{Landau:2}. Examples are the longitudinal and synchronous  gauges. In the FRW universe this 
condition is fulfilled automatically.}   $\pi=g^{00}\sqrt{-g}\,\nabla_0\varphi$, 
we find for the derivative of the spectral function 
\begin{align}
\label{Grhoder}
\nabla^x_0G_\rho(x,y)=\frac{\delta(\vec{x},\vec{y}\,)}{g^{00}\sqrt{-g}}\,.
\end{align}
Multiplication of \eqref{Grhoder} by $g^{00}\delta(x^0,y^0)$ 
then  gives the generalized $\delta$ function $\delta^g(x,y)$,
which cancels the generalized $\delta$ function on 
the right--hand side of \eqref{convolution}.

The local term of the self--energy (\ref{selfenergy}), proportional 
to the $\delta$ function, can be absorbed in the effective mass 
\begin{align}
M^2(x)\equiv M^2+{\frac{\lambda}{2}}G(x,x)\,.
\end{align}
The remaining part of the self--energy can also be split 
into a spectral part, $\Pi_\rho(x,y)$, and a statistical 
part,  $\Pi_F(x,y)$, in complete analogy to \eqref{GfGr}.

Integrating along the closed time path in the direction 
indicated in Fig.\,\ref{diagrams}, and taking into account 
that any point of the negative branch is considered as a 
later instant than any point of the positive branch, we 
finally obtain the system of Kadanoff--Baym equations:
\begin{widetext}
\begin{subequations}
\label{KBeqs}
\begin{align}
\label{GFeq}
[\square_x&+M^2(x)]G_F(x,y)=
{\int\limits^{y^0}_0}\sqrt{-g}d^4z\,\Pi_F(x,z)G_\rho(z,y)
-{\int\limits^{x^0}_0}\sqrt{-g}d^4z\,\Pi_\rho(x,z)G_F(z,y)\,,\\
\label{Grhoeq}
[\square_x&+M^2(x)]G_\rho(x,y)=-
{\int\limits^{x^0}_{y^0}}\sqrt{-g}d^4z\,\Pi_\rho(x,z)G_\rho(z,y)\,.
\end{align}
\end{subequations}
\end{widetext}
Comparing with the Kadanoff--Baym equations presented in 
\cite{Berges:2001fi,Lindner:2005kv}, we conclude that \eqref{KBeqs} appear
to be the covariant generalization of the Kadanoff--Baym equations 
in  Minkowski space--time. 

Equations \eqref{KBeqs} are exact equations for the quantum 
dynamical evolution of the  statistical propagator and spectral 
function. It is important that, due to the characteristic memory 
integrals on the right--hand sides, the dynamics of the system 
depends on the history of its evolution \cite{Berges:2005vj}.

To complete this section we  derive explicit expressions for 
the spectral and statistical self--energies. Using symmetry 
(antisymmetry) of the spectral and statistical propagators
with respect to permutation of the arguments, we obtain for 
the three--loop contribution to the self--energy components:
\begin{subequations}
\label{PiFrho3loop}
\begin{align}
\Pi^{(3)}_F(x,y)=&-\frac{\lambda^2}{6}[G_F(x,y)G_F(x,y)G_F(x,y)\nonumber\\
&-{\textstyle\frac34} G_F(x,y)G_\rho(x,y)G_\rho(x,y)]\,,\\
\Pi^{(3)}_\rho(x,y)=&-\frac{\lambda^2}{6}[3G_F(x,y)G_F(x,y)G_\rho(x,y)\nonumber\\
&-{\textstyle\frac14} G_\rho(x,y)G_\rho(x,y)G_\rho(x,y)]\,.
\end{align}
\end{subequations}
Four-- and higher--loop contributions to the self--energy 
components contain integrations over  space--time  with $x^0$ 
and $y^0$ as the integration limits. Introducing
\begin{subequations}
\label{G4FG4Rhodef}
\begin{align}
G_{4F}(x,y)&=\txtint_0^{x^0}
\sqrt{-g}d^4z\, G_F(x,z)G_\rho(x,z)  \nonumber\\
&\times [G^2_F(z,y)-{\textstyle\frac14}G^2_\rho(z,y)]+\{x\leftrightarrow y\}\,,\\
G_{4\rho}(x,y)&=\txtint_{0}^{x^0} 
\sqrt{-g}d^4z\, G_F(x,z)G_\rho(x,z) \nonumber\\
&\times [2G_F(z,y)G_\rho(z,y)]-\{x\leftrightarrow y\}\,,
\end{align}
\end{subequations}
we can write the four--loop contribution to the statistical 
and spectral components of the self--energy as
\begin{subequations}
\label{PiFrho4loop}
\begin{align}
\Pi^{(4)}_F(x,y)=&\frac{\lambda^3}{2}[G_F(x,y)G_{4F}(x,y)\nonumber\\
&\hspace{18mm}-{\textstyle\frac14}G_\rho(x,y)G_{4\rho}(x,y)]\,,\\
\Pi^{(4)}_\rho(x,y)=&\frac{\lambda^3}{2}[G_F(x,y)G_{4\rho}(x,y)\nonumber\\
&\hspace{18mm}+G_\rho(x,y)G_{4F}(x,y)]\,.
\end{align}
\end{subequations}

Of course, all  quantities entering the Kadanoff--Baym equations
must be renormalized. The renormalization  
at finite temperature has been developed in 
\cite{vanHees:2001ik,Blaizot:2003br,Berges:2005hc,Arrizabalaga:2005tf}. 
A generalization to out--of--equilibrium systems with non--Gaussian 
initial conditions has been obtained in \cite{Borsanyi:2008ar,Garny:2009ni}.
A renormalization procedure at  tadpole order in the  Gaussian scheme 
in the expanding universe has been applied to the analysis of  
Kadanoff--Baym equations in \cite{Tranberg:2008ae}.

\section{\label{QK}Quantum kinetics}

Introducing the retarded and advanced propagators 
\begin{subequations}
\label{GRGAdef}
\begin{align}
G_R(x,y)&\equiv\theta(x^0-y^0)G_\rho(x,y)\,,\\
G_A(x,y)&\equiv-\theta(y^0-x^0)G_\rho(x,y)\,,
\end{align}
\end{subequations} 
and the corresponding definitions for the self--energies, 
one can rewrite the system of Kadanoff--Baym equations in 
the form:
\begin{subequations}
\label{KBQK}
\begin{align}
\label{GFQK1}
[\square_x&+M^2(x)]G_F(x,y)=-\txtint
\sqrt{-g}d^4z
\theta(z^0)\nonumber\\
&\times[\Pi_F(x,z)G_A(z,y)+\Pi_R(x,z)G_F(z,y)]\,,\\
\label{GrhoQK1}
[\square_x&+M^2(x)]G_\rho(x,y)=-\txtint
\sqrt{-g}d^4z\theta(z^0)\nonumber\\
&\times[\Pi_\rho(x,z)G_A(z,y)+\Pi_R(x,z)G_\rho(z,y)]\,.
\end{align}
\end{subequations}
The system \eqref{KBQK} should be supplemented by the analogous 
equations for the retarded (advanced) propagators; they can be 
derived from \eqref{Grhoeq} upon use of \eqref{Grhoder}
\begin{align}
\label{GRKQ1}
\big[\square_x+M^2&(x)\big]G_{R(A)}(x,y)=\delta^g(x,y)\nonumber\\
&-\txtint\sqrt{-g}d^4z\,\Pi_{R(A)}(x,z)G_{R(A)}(z,y)\,.
\end{align}
\newpage
Let us now interchange $x$ and $y$ on both sides of the 
Kadanoff--Baym equations \eqref{KBQK}. The difference
(sum) of the original and resulting equations are referred to as
the kinetic (constraint) equations for the spectral 
function and the statistical propagator. Using the relation $G_R(x,y)=G_A(y,x)$ 
and symmetry (antisymmetry) of the statistical propagator (spectral function) 
we obtain

\begin{widetext}
\begin{subequations}
\label{GFGrhodifference}
\begin{align}
\label{GFdifference}
&[\square_x\mp\square_y+M^2(x)\mp M^2(y)]G_F(x,y)=\nonumber\\&\hspace{20mm}=-\txtint
\sqrt{-g} d^4z\theta(z^0) [\Pi_F(x,z)G_A(z,y)\mp G_R(x,z)\Pi_F(z,y)\nonumber\\&\hspace{55mm}
+\Pi_R(x,z)G_F(z,y)\mp G_F(x,z)\Pi_A(z,y)]\,,\\
\label{Grhodifference}
&[\square_x\mp\square_y+M^2(x)\mp M^2(y)]G_\rho(x,y)=\nonumber\\&\hspace{20mm}=-\txtint
\sqrt{-g} d^4z \theta(z^0) [\Pi_\rho(x,z)G_A(z,y)\mp G_R(x,z)\Pi_\rho(z,y)\nonumber\\&\hspace{55mm}
+\Pi_R(x,z)G_\rho(z,y)\mp G_\rho(x,z)\Pi_A(z,y)]\,,
\end{align}
\end{subequations}
Interchanging $x$ and $y$ on both sides of the equation for 
$G_A(x,y)$ and adding it to the equation for $G_R(x,y)$
we obtain the constraint equation for the retarded propagator:
\begin{align}
\label{GRsum}
&[\square_x+\square_y+M^2(x)+M^2(y)]G_R(x,y)=\nonumber\\&\hspace{20mm}=2\delta^g(x,y)
-\txtint \sqrt{-g} d^4z  [\Pi_R(x,z)G_R(z,y)+G_R(x,z)\Pi_R(z,y)]\,.
\end{align}
\end{widetext}

Next, we introduce center and relative coordinates. In 
Minkowski space--time they are given by half of the sum and 
by the difference of $x$ and $y$, respectively \cite{Lindner:2005kv}.
In other words the center coordinate lies in the middle of 
the geodesic  connecting $x$ and $y$, whereas the relative 
coordinate gives the length of the ``curve''\footnote{In 
Minkowski space--time  geodesics are straight lines.} 
connecting the two points.

Consider now  curved space--time.
\begin{figure}[!ht]
\begin{center}
\includegraphics[width=0.4\textwidth]{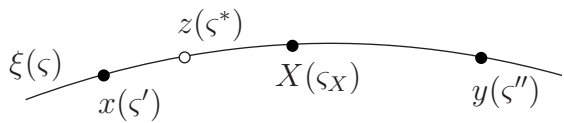}
\end{center}
\caption{\label{geodesic}Arrangement of the points along the geodesic.}
\end{figure} 
Let $\varsigma$ be the affine parameter of the geodesic connecting 
$x$ and $y$ (see Fig.\,\ref{geodesic}) and $\xi(\varsigma)$ a function 
mapping $\varsigma$ onto the points of the geodesic, with
\begin{align}
x^\alpha=\xi(\varsigma'),\quad y^\alpha=\xi(\varsigma'')\,.
\end{align}
The center coordinate lies in the middle of the geodesic,
i.e. it corresponds to $\varsigma_X\equiv\frac12(\varsigma'+\varsigma'')$. 
The relative coordinate is given by the sum of the infinitesimal
distance vectors $d\xi^\alpha$ along the geodesic, all of which 
must have been submitted to parallel transfer to $\varsigma_X$ from the 
integration point on the curve\footnote{
Calzetta and Hu \cite{Calzetta:1987,Calzetta:1987bw} have employed a different 
method based on the use of Riemann normal coordinates and the momentum 
representation of the propagators. Their approach has some advantages 
for the study of the quantum kinetics equations. Here we are mainly
interested in the Kadanoff--Baym and Boltzmann equations and consider the 
derivation of the quantum kinetic equations as an intermediate
step connecting both of them. For this reason, we adopt the covariant 
definitions of the midpoint and distance vectors introduced by 
Winter \cite{PhysRevD.32.1871}, which allow us to keep the analysis 
 manifestly covariant in every step.
}. According to 
\cite{PhysRevD.32.1871} this implies
\begin{align}
X^\alpha\equiv X^\alpha_{xy}=\xi^\alpha\left(\varsigma_X\right),~
s^\alpha\equiv s^\alpha_{xy}=(\varsigma'-\varsigma'')u^\alpha\left(\varsigma_X\right)\,.
\end{align}

All quantities in equations \eqref{GFGrhodifference} are now   
recast in terms of $X^\alpha$ and $s^\alpha$. Up to higher order,
proportional to the curvature tensor terms, the Laplace--Beltrami 
operator is given by \cite{PhysRevD.32.1871}
\begin{align}
\label{DAlambertXs}
\square_{x,y}\approx{\frac14}D^\alpha D_\alpha+
\frac{\partial^2}{\partial s^\alpha\partial s_\alpha}
\pm D^\alpha\frac{\partial}{\partial s_\alpha}\,,
\end{align}
where $D_\alpha$ is the covariant derivative
\begin{align}
D_\alpha&\equiv \frac{\partial }{\partial X^\alpha}-
\Gamma^\beta_{\alpha\gamma}s^\gamma\frac{\partial}{\partial s^\beta}\,.
\end{align}
Note that in \eqref{DAlambertXs} we have neglected the corrections 
proportional to the Riemann and Ricci tensors. 
Next, we Taylor expand the effective masses to first order 
around the center coordinate $X$
\begin{align}
\label{MTaylor}
M^2\approx M^2(X)\pm{\textstyle\frac12}s^\alpha D_\alpha M^2(X)\,,
\end{align}
where the minus sign corresponds to $y$ whereas the plus sign
corresponds to $x$.
The propagators on the left--hand side of \eqref{GFGrhodifference} 
can also be reparameterized in terms of the center and relative 
coordinates: $G_F(x,y)\rightarrow \tilde{G}_F(X,s)$ and 
$G_\rho(x,y)\rightarrow \tilde{G}_\rho(X,s)$.

On the right--hand sides we have convolutions of functions of 
$x$ and $z$ and functions of $z$ and $y$. That is, we have to 
introduce the corresponding center and relative coordinates and
perform the integration. Making use of the identity 
\[
(\varsigma'+\varsigma^*)=(\varsigma'+\varsigma'')+
(\varsigma^*-\varsigma'')=2\,\varsigma_X+(\varsigma^*-\varsigma'')
\]
and Taylor expanding around $\varsigma_X$, we obtain to first order
\begin{align}
\label{PiFexpansion}
\Pi_F(x,z)&\equiv \tilde{\Pi}_F(X_{xz},s_{xz})\approx 
\tilde{\Pi}_F(X,s_{xz})\nonumber\\
&+\left({\frac{\partial \tilde{\Pi}_F}{\partial \xi^\alpha}}
\frac{d\xi^\alpha}{d\varsigma}
+{\frac{\partial \tilde{\Pi}_F}{\partial u^\alpha}}
\frac{du^\alpha}{d\varsigma}\right)\frac{\varsigma^*-\varsigma''}2\,.
\end{align}
Using furthermore the definition of the four--velocity and the 
geodesic equation 
\begin{align}
\frac{d\xi^\alpha}{d\varsigma}=u^\alpha,\quad 
\frac{du^\alpha}{d\varsigma}=-\Gamma^\alpha_{\beta\gamma}u^\beta u^\gamma\,,
\end{align}
we can rewrite \eqref{PiFexpansion} in the form 
\begin{align}
\label{PiFxz}
\Pi_F(x,z)\approx \tilde{\Pi}_F(X,s_{xz})+
{\textstyle\frac12} s^\alpha_{zy}D_\alpha \tilde{\Pi}_F(X,s_{xz})\,,
\end{align}
where $s^\alpha_{zy}\equiv (\varsigma^*-\varsigma'')u^\alpha(\varsigma_X)$. 
Making use of the identity 
\[
(\varsigma''+\varsigma^*)=(\varsigma'+\varsigma'')-
(\varsigma'-\varsigma^*)=2\,\varsigma_X-(\varsigma'-\varsigma^*)
\]
we get a similar expression for the functions of $z$ and $y$
\begin{align}
\label{GAzy}
G_A(z,y)\approx \tilde{G}_A(X,s_{zy})-
{\textstyle\frac12} s^\alpha_{xz}D_\alpha \tilde{G}_A(X,s_{zy})\,.
\end{align}
To perform the integration of the product of \eqref{PiFxz} and 
\eqref{GAzy}, we shift the coordinate origin to $\varsigma_X$ and replace 
the integration with respect to $z$ by integration with respect
to distance $s_{Xz}$ from $X$ to $z$ along the geodesic. Moreover, 
we approximate\footnote{The next--to--leading term of the Taylor
expansion is proportional to the convolution of the Christoffel 
symbol \cite{Landau:2}, $\sqrt{-g}_z\approx\sqrt{-g}_X(1+
\Gamma^\nu_{\alpha\nu}s^\alpha)$. This correction can in principle be 
taken into account and would induce {} additional terms proportional to 
$i\partial/\partial p^\alpha$ on the right--hand side of the 
quantum kinetic equation. Since such term are neglected
in the Boltzmann approximation,  the collision 
terms do not receive any corrections.} $\sqrt{-g}_z$ by its value 
at the origin $\sqrt{-g}_X$.

The Kadanoff--Baym equations describe the dynamics of a system in 
terms of the spectral function and statistical propagator. The 
latter ones are functions of two coordinates in the four--dimensional 
space--time. By introducing  center and relative coordinates 
we have traded one set of coordinates for another one. 
Performing the so--called Wigner transformation, one can also trade one 
of the arguments defined in the coordinate space for an argument defined 
in the momentum space. In  curved space--time \cite{PhysRevD.32.1871}
\begin{subequations}
\label{WignerTrafos}
\begin{align}
\label{GFXp}
\tilde{G}_F(X,p)&=\sqrt{-g}_X\int d^4s\, e^{ips}\tilde{G}_F(X,s)\,,\\
\label{GFXs}
\tilde{G}_F(X,s)&=\frac{1}{\sqrt{-g}_X}\int \frac{d^4p}{(2\pi)^4}\,e^{-ips}\tilde{G}_F(X,p)\,.
\end{align}
\end{subequations}
Note that in \eqref{WignerTrafos} and in the rest of the paper we 
use \textit{contravariant} components of the space--time coordinates 
and \textit{covariant} components of the momenta. Let us also note  
that
\begin{align*}
d\Pi^4_p\equiv \frac1{\sqrt{-g}_X}\frac{d^4p}{(2\pi)^4}
\end{align*}
is the invariant volume element in  momentum space. The definition 
of the Wigner transform of $\tilde{G}_\rho(X,s)$ differs from \eqref{GFXp} 
by a factor of $-i$ so that $\tilde{G}_\rho(X,p)$ is again real valued.

As follows from \eqref{GFXs}, differentiation with respect to $s^\alpha$
is replaced after the Wigner transformation by $p^\alpha$
\begin{align}
\frac{\partial}{\partial s^\alpha}\rightarrow -ip^\alpha\,.
\end{align}
Upon integration by parts we also  see that $s^\alpha$ is replaced
by differentiation with respect to
$p^\alpha$:
\begin{align}
\label{s_rule}
s^\alpha\rightarrow -i\frac{\partial}{\partial p^\alpha}\,.
\end{align}
Consequently the Wigner transformed covariant derivative reads
\begin{align}
\label{CovDer}
D_\alpha\rightarrow {\cal D}_\alpha=\frac{\partial}{\partial X^\alpha}
+\Gamma^\beta_{\alpha\gamma}p_\beta\frac{\partial}{\partial p_\gamma}\,.
\end{align}

Correlations between earlier and later times are exponentially suppressed, 
which leads to a gradual loss of the dependence
on the initial conditions \cite{Berges:2005vj,Lindner:2005kv}. 
Exploiting this fact, one can drop the $\theta$ function from 
the integrals in the difference equations \eqref{GFGrhodifference}. 
Furthermore we let the relative--time coordinate $s^0$  range
from $-\infty$ to $\infty$ in order to perform the Wigner transformation,
see \cite{Berges:2005vj,Berges:2005md} for a detailed discussion 
of these approximations.
Then using  \eqref{s_rule} and \eqref{CovDer} we obtain for the 
Wigner transform of the first term on the right--hand side 
of \eqref{GFdifference}:
\begin{align}
\label{WTofConvolution}
\hspace{-1mm}
\txtint & \sqrt{-g}_zd^4z~ \Pi_F(x,z)G_A(z,y)\rightarrow \nonumber\\
&\tilde{\Pi}_F(X,p)\tilde{G}_A(X,p)
+{\textstyle\frac{i}{2}}\{ 
\tilde{\Pi}_F(X,p),\tilde{G}_A(X,p)
\}_{PB}\,,
\end{align}
where the Poisson brackets are defined by 
\begin{align}
\label{PoissonBrackets}
\{\tilde{A}(X,p),\tilde{B}(X,p)\}_{PB}&\equiv
\frac{\partial}{\partial p_\alpha} \tilde{A}(X,p) {\cal D}_\alpha\tilde{B}(X,p)\nonumber\\
&-{\cal D}_\alpha \tilde{A}(X,p) \frac{\partial}{\partial p_\alpha}\tilde{B}(X,p)\,.
\end{align}
Comparing \eqref{PoissonBrackets} to its Minkowski--space counterpart 
we see that the derivatives with respect to $X$ are replaced by the 
covariant derivatives, just as one would expect.

Wigner transforming the rest of the terms we obtain a rather lengthy 
expression which can be substantially simplified with the help of the 
relations between $\tilde{G}_R(X,p)$, $\tilde{G}_A(X,p)$, and 
$\tilde{G}_\rho(X,p)$. Recalling the Fourier transform of the 
$\theta$ function,
\[
\txtint ds^0\exp(iws^0)\theta(\pm s^0)=\lim\limits_{\epsilon\rightarrow 0}
\frac{\pm\, i}{\omega\pm i\epsilon},
\]
we find that
\begin{subequations}
\label{GAGR}
\begin{align}
\label{GRexpr}
\tilde{G}_R(X,p)&=-\int\frac{d\omega}{2\pi}\frac{\tilde{G}_\rho(X,\vec{p},\omega)}{p_0-\omega+i\epsilon}\,,\\
\label{GAexpr}
\tilde{G}_A(X,p)&=-\int\frac{d\omega}{2\pi}\frac{\tilde{G}_\rho(X,\vec{p},\omega)}{p_0-\omega-i\epsilon}\,.
\end{align}
\end{subequations}
From comparison of \eqref{GAexpr} and \eqref{GRexpr} it follows that
\begin{align}
\label{GRGAcc}
\tilde{G}_A(X,p)=\tilde{G}^*_R(X,p)\,.
\end{align}
Recalling furthermore that the $\delta$ function can be approximated by 
\begin{align}
\label{deltalimit}
\delta(\omega)=\lim\limits_{\epsilon\rightarrow +0}\frac{\epsilon}{\pi(\omega^2+\epsilon^2)}\,,
\end{align}
we also find that 
\begin{align}
\label{GRminGA}
\tilde{G}_R(X,p)-\tilde{G}_A(X,p)=i\tilde{G}_\rho(X,p)\,.
\end{align}
Analogous relations also hold for the retarded and advanced 
components of the self--energy.

As can be inferred from \eqref{DAlambertXs} and \eqref{MTaylor}, the 
Wigner transform of the left--hand side of \eqref{GFGrhodifference} 
reads \footnote{Additional contributions arising from the decomposition 
of the Laplace--Beltrami operator are proportional to  Riemann and 
Ricci tensors and  to the curvature (see Eq. (4.40) in \cite{PhysRevD.32.1871}) 
and may be relevant in strong  gravitational fields. Since all these terms contain 
at least one $i\partial/\partial p_\alpha$ derivative, they do not 
contribute in the Boltzmann approximation.}
\begin{align}
\square_x-\square_y+M^2(x)&-M^2(y)\rightarrow\nonumber\\
&-i\left(2p^\alpha{\cal D}_\alpha+D_{\alpha}M^2\frac{\partial}{\partial p_\alpha}\right)\,.
\end{align}
Introducing the quantity 
\begin{align}
\label{Omega}
\tilde{\Omega}(X,p)\equiv p^\mu p_\mu -M^2(X)-\tilde{\Pi}_h(X,p)\,,
\end{align}
where $\tilde{\Pi}_h(X,p)\equiv {\rm Re}\,\tilde{\Pi}_R(X,p)$,
and collecting the terms on the right--hand side of the kinetic 
equation \eqref{GFGrhodifference} one can write the kinetic equation for the 
Wigner transform of the statistical propagator in the compact form:
\begin{align}
\label{GFQKequation}
\{\tilde{\Omega}(X,p),\tilde{G}_F&(X,p)\}_{PB}\nonumber\\
&=\tilde{G}_F(X,p)\tilde{\Pi}_\rho(X,p)-\tilde{\Pi}_F(X,p)\tilde{G}_\rho(X,p)\nonumber\\
&+\{\tilde{\Pi}_F(X,p),\tilde{G}_h(X,p)\}_{PB}\,,
\end{align}
where $\tilde{G}_h(X,p)\equiv {\rm Re}\,\tilde{G}_R(X,p)$.
The same procedure leads also to a kinetic equation for
the Wigner transform of the spectral function
\begin{align}
\label{GrhoQKequation}
\{\tilde{\Omega}(X,p),\tilde{G}_\rho&(X,p)\}_{PB}\nonumber\\
&=\{\tilde{\Pi}_\rho(X,p),\tilde{G}_h(X,p)\}_{PB}\,.
\end{align}

As has been mentioned in the previous section, the exact quantum 
dynamical evolution of the system depends on its whole evolution history.
Mathematically, this manifests itself in the memory integrals 
on the right--hand sides of \eqref{KBeqs}. In fact, performing the 
linear order Taylor expansion around $X$, we  take into 
account only a very short part of the history of the evolution. Since 
the expansion coefficients are defined at $X$, after the integration 
we obtain equations which are \textit{local} in time.

Next we consider the Wigner transform of the constraint equation for the 
retarded propagator \eqref{GRsum}. On the 
left--hand side we have $\square_x+\square_y=2\partial_{s^\alpha}\partial_{s_\alpha}$,
to first order in the covariant derivative,
 whereas $M^2(x)+M^2(y)\approx M^2(X)$. On the right--hand side
the Poisson brackets cancel out and only the product of $\tilde{\Pi}_R(X,p)$
and $\tilde{G}_R(X,p)$ remains. Finally, the  Wigner transform of the generalized
$\delta$ function is just unity. Therefore, we get an 
\textit{algebraic} equation for the Wigner transform of the retarded
propagator
\begin{align}
\label{GRsolution}
[\,p^\mu p_\mu -M^2(X)-\tilde{\Pi}_R(X,p)\,]\tilde{G}_R(X,p)=-1\,.
\end{align}
Equation \eqref{GRsolution} implies that the real 
part of the retarded propagator is given by
\begin{align}
\label{GRsol}
\tilde{G}_h(X,p)=\frac{-\tilde{\Omega}(X,p)}{\tilde{\Omega}^2(X,p)+
{\textstyle \frac{1}4}\tilde{\Pi}^2_\rho(X,p)}\,.
\end{align}
Note that $\tilde{G}_h(X,p)$ vanishes on the mass shell, which is 
defined by the condition $\tilde{\Omega}(X,p)=0$. As follows from 
\eqref{GRGAcc} and  \eqref{GRminGA}, the Wigner transform of the 
spectral function is twice the imaginary part of the retarded 
propagator:
\begin{align}
\label{Grhosol}
\tilde{G}_\rho(X,p)=\frac{-\tilde{\Pi}_\rho(X,p)}{\tilde{\Omega}^2(X,p)+
{\textstyle\frac14}\tilde{\Pi}^2_\rho(X,p)}\,.
\end{align}

Equation \eqref{Grhosol} is also a solution of \eqref{GrhoQKequation}.
To first order in the covariant derivative the Wigner--transform of the 
constraint equation for the statistical propagator reads
\begin{align}
\label{GFconstraint}
&\tilde{\Omega}(X,p)\tilde{G}_F(X,p)={\textstyle\frac14}\{\tilde{\Pi}_F(X,p),
\tilde{G}_\rho(X,p)\}_{PB}\nonumber\\
&+{\textstyle\frac14}\{\tilde{G}_F(X,p),\tilde{\Pi}_\rho(X,p)\}_{PB}
+\tilde{\Pi}_F(X,p)\tilde{G}_h(X,p).
\end{align}
The constraint equation for $\tilde{G}_F(X,p)$ is no longer algebraic 
and can not be solved analytically in general. However,
let us assume for a moment that the system is in thermal equilibrium. 
In this case all the quantities are constant in time and space and 
the Poisson brackets in \eqref{GFconstraint} vanish identically. 
The solution of the resulting algebraic equation then reads
\begin{align}
\label{GFequilibrium}
\tilde{G}^{eq}_{F}(p)=
\frac{\tilde{\Pi}_{F}(p)}{\tilde{\Pi}_{\rho}(p)}\, \tilde{G}^{eq}_{\rho}(p)\,.
\end{align}
That is, we have obtained the fluctuation--dissipation relation. 
It only remains to calculate the ratio of the spectral and statistical 
self--energies. This can be done using the relation \eqref{GfGr}
and the KMS periodicity condition, $G(x,y)|_{x=0}=G(x,y)|_{x=-i\beta}$,
where $\beta$ is the inverse temperature. Wigner--transforming this
equation and using \eqref{GFequilibrium} we obtain
\begin{align}
\label{fluct_diss}
\tilde{G}^{(eq)}_F(p)=\left[
n^{(eq)}(p)+{\textstyle\frac12}\right]
\tilde{G}^{(eq)}_\rho(p)\,,
\end{align}
where $n^{(eq)}$ is the Bose--Einstein distribution function.

To complete this section, we have to express the Wigner transforms 
of the spectral and statistical self--energies in terms of the 
Wigner transforms of the spectral function and statistical propagator.
Using the definitions of the Wigner transformation and its 
inverse we find for the Wigner transform of a product of functions 
of the same arguments:
\begin{align}
f_1&(x,y)\ldots f_n(x,y)\rightarrow \widetilde{f_1\ldots f_n}(X,p)\nonumber\\
&\equiv \txtint d\Pi^4_{p_1} \ldots d\Pi^4_{p_n}
 (2\pi)^4 \sqrt{-g}_X \delta^4 (-p+p_1+\ldots p_n)\nonumber\\
&\hspace{35mm}\times\tilde{f}(X,p_1)\ldots \tilde{f}(X,p_n)\,.
\end{align}
Note that $\delta_g(q)\equiv \sqrt{-g}_X\, \delta(q)$ represents the momentum--space 
generalization of the $\delta$ function, invariant under  coordinate 
transformations (this can be checked with help of the scaling property 
of the $\delta$ function). Keeping in mind that the definition 
of $\tilde{G}_\rho(X,p)$ contains an additional factor of $-i$ we 
can then  write  the Wigner transforms of \eqref{PiFrho3loop} in the form
\begin{subequations}
\label{PiFrho3loopWigner}
\begin{align}
\tilde{\Pi}^{(3)}_F(X,p)=&-\frac{\lambda^2}{6}[\,\widetilde{G^3_F}(X,p)
+{\textstyle\frac34} \widetilde{G_FG^2_\rho}(X,p)]\,,\\
\tilde{\Pi}^{(3)}_\rho(X,p)=&-\frac{\lambda^2}{6}[3\widetilde{G^2_FG_\rho}(X,p)
+{\textstyle\frac14} \widetilde{G^3_\rho}(X,p)]\,.
\end{align}
\end{subequations}
The expression for the Wigner transform of the three--loop 
retarded self--energy can be obtained from \eqref{PiFrho3loopWigner} by 
replacing one of the $\tilde{G}_\rho$ by $\tilde{G}_R$. 
The Wigner transforms 
of the four--loop contributions \eqref{PiFrho3loop} can be written in a 
similar way 
\begin{subequations}
\label{PiFrho4loopWigner}
\begin{align}
\tilde{\Pi}^{(4)}_F(X,p)=&\frac{\lambda^3}{2}[\widetilde{G_FG_{4F}}(X,p)+
{\textstyle\frac14}\widetilde{G_\rho G_{4\rho}}(X,p)]\,,\\
\label{Pi4rho}
\tilde{\Pi}^{(4)}_\rho(X,p)=&\frac{\lambda^3}{2}[\widetilde{G_FG_{4\rho}}(X,p)
+\widetilde{G_\rho G_{4F}}(X,p)]\,.
\end{align}
\end{subequations}
Note, however, that $\tilde{G}_{4F}$ and $\tilde{G}_{4\rho}$ 
are Wigner transforms of \textit{convolutions} of four two--point 
functions, 
\begin{subequations}
\begin{align}
G_{4F}(x,y)&=\txtint 
\sqrt{-g}d^4z\, G_F(x,z)G_R(x,z)  \nonumber\\
&\times [G^2_F(z,y)-{\textstyle\frac14}G^2_\rho(z,y)]+\{x\leftrightarrow y\}\,,\\
G_{4\rho}(x,y)&=\txtint 
\sqrt{-g}d^4z\, G_F(x,z)G_R(x,z) \nonumber\\
&\times [2G_F(z,y)G_\rho(z,y)]-\{x\leftrightarrow y\}\,,
\end{align}
\end{subequations}
where we have used the definitions of the retarded and advanced 
propagators and dropped again the $\theta(z^0)$ factor. Proceeding
as in Eq.\,\eqref{WTofConvolution} and making use of the relations 
\eqref{GRGAcc} and \eqref{GRminGA}, we obtain for the Wigner transforms
of $G_{4F}$ and $G_{4\rho}$
\begin{subequations}
\begin{align}
&\tilde{G}_{4F}(X,p)=2
[\widetilde{G^2_F}(X,p)+{\textstyle\frac14}\widetilde{G^2_\rho}(X,p)]
\widetilde{G_F\Re[G_R]}(X,p)\nonumber\\
&+{\textstyle\frac12}\{\widetilde{G^2_F}(X,p)+
{\textstyle\frac14}\widetilde{G^2_\rho}(X,p),\widetilde{G_FG_\rho}(X,p)\}_{P.B.}\,,\\
&\tilde{G}_{4\rho}(X,p)=4\widetilde{G_FG_\rho}(X,p)\widetilde{G_F\Re[G_R]}(X,p)\,.
\end{align}
\end{subequations}
Finally, the expression for the Wigner transform of the 
four--loop retarded self--energy can be obtained 
from \eqref{Pi4rho} by replacing $\tilde{G}_\rho$ with $\tilde{G}_R$ 
and $\tilde{G}_{4\rho}$ with $\tilde{G}_{4R}$. The latter one is related 
to $\tilde{G}_{4\rho}$ by Eq.\,\eqref{GRexpr}.

\section{\label{BE}Boltzmann kinetics}

The spectral function \eqref{Grhosol} has approximately  Breit--Wigner 
shape with a width proportional to the spectral self--energy. The area 
under $G_\rho(X,p)$ is determined by the normalization condition, 
\begin{align}
\label{Grhonormalization}
\int \frac{ g^{00}}{2\pi}  \tilde{G}_\rho(X,p)\, p_0\, dp_0=1,
\end{align}
which is a direct consequence of \eqref{Grhoder} and the antisymmetry 
of the spectral function with respect to permutation of its arguments.
In the limit of vanishing coupling constant the width of the spectral 
function approaches zero, whereas its on--shell value goes to infinity, 
see Eq.\,\eqref{Grhosol}. Equation \eqref{deltalimit} then implies that
in this limit the spectral function takes the quasiparticle form \cite{KB:1962}
\begin{align}
\label{QPapprox}
\tilde{G}_\rho(X,p)=2\pi\,\sign(p_0) \, \delta\left(g^{\mu\nu} p_\mu p_\nu-M^2\right)\,.
\end{align}
Note that \eqref{QPapprox} is consistent with the normalization condition
\eqref{Grhonormalization}. The signum--function appears in \eqref{QPapprox} 
because $\tilde{\Pi}_\rho(X,p)$ is an odd function of $p_0$. Since the magnitudes 
of $\tilde{\Pi}_\rho$,  $\tilde{\Pi}_h$ and of the local term of the 
self--energy are controlled by the same coupling we have also neglected them
in $\tilde{\Omega}(X,p)$.
In the same limit Eq.\,\eqref{GrhoQKequation} for the spectral function 
simplifies to 
\begin{align}
\label{BEGrho}
p^\alpha {\cal D}_\alpha \tilde{G}_\rho(X,p)&=0
\end{align}
and indeed admits a quasiparticle solution \eqref{QPapprox}. Note that 
Eqs.\,\eqref{BEGrho} and \eqref{QPapprox} state that the effective mass 
$M$ of the field quanta does not change as they move along the geodesic, 
just like it is the case for particles.

Motivated by the fluctuation--dissipation relation \eqref{fluct_diss}
we can trade the statistical propagator for some other function:
\begin{align}
\label{KBAnsatz}
\tilde{G}_F(X,p)=\left[n(X,p)+{\textstyle\frac12}\right]
\tilde{G}_\rho(X,p)\,.
\end{align}
However, if both $\tilde{G}_F(X,p)$ and $\tilde{G}_\rho(X,p)$ are smooth functions
then relation \eqref{KBAnsatz} is merely a definition of $n(X,p)$. In the 
quasiparticle approximation the spectral function is divergent 
and forces the momentum argument of $n$ to be on the mass shell. For this 
reason the quasiparticle approximation for the statistical propagator 
\eqref{KBAnsatz} is usually referred to as the Kadanoff--Baym \textit{Ansatz} 
\cite{KB:1962,Berges:2004yj}.

Let us now tentatively put the coupling constant to zero. In this case 
the right--hand sides 
of the kinetic equations \eqref{GFQKequation} and \eqref{GrhoQKequation}
vanish. In this case $\tilde{G}_F(X,p)$ and $\tilde{G}_\rho(X,p)$ are 
constant in space and time even if the system is out of equilibrium. If 
we now ``increase'' the coupling constant again, then the gain and loss terms 
on the right--hand side of \eqref{GFQKequation} will induce nontrivial
dynamics for the statistical propagator. This in turn will induce a time and 
space dependence of the spectral and statistical propagators thus leading 
to nonvanishing Poisson brackets on the right--hand sides of the kinetic
equations. The magnitude of the derivatives with respect to the time and
space coordinates are therefore proportional to some (positive) power of the coupling
constant. Consequently the contribution of the Poisson brackets in 
\eqref{GFQKequation} is effectively of higher order in $\lambda$ than 
the contribution of the gain and loss terms. These considerations justify the dropping of 
the Poisson brackets and of the local and nonlocal contributions
to the effective field mass in the kinetic equations. In other words, they 
legitimate the use of the quasiparticle approximation.\footnote{If we were 
interested in higher order processes, for instance in the $2\rightarrow 4$
scattering which is of the \textit{fourth} order in the coupling constant, 
we would have to use the so called \textit{extended quasiparticle approximation}
\cite{PhysRevB.52.14615,PhysRevC.46.1687,PhysRevC.48.1034,PhysRevC.64.024613,
CondMatPhys2006_9_473,JPhys2006_35_110}. Since in this paper we limit ourselves 
to the processes of at most \textit{third} order in $\lambda$, the quasiparticle
approximation is sufficient for our purposes.}

As has been argued above, the Poisson brackets partially take into 
account the memory effects. Neglecting the Poisson brackets we 
completely ignore the previous evolution of the system. Physically 
this corresponds to the \textit{Stosszahlansatz} of Boltzmann.

From Eqs.\,\eqref{GFQKequation}, \eqref{BEGrho} and \eqref{KBAnsatz} it 
follows that in this approximation the kinetic equation for the 
statistical propagator turns into an equation for the evolution of the 
one--particle distribution function $n(X,p)$:
\begin{align}
\label{BoltzmannPreliminary}
[\,p^\alpha &{\cal D}_\alpha n(X,p)]\tilde{G}_\rho(X,p)\nonumber\\
&={\textstyle\frac12}[\tilde{\Pi}_{>}(X,p)\tilde{G}_{<}(X,p)
-\tilde{G}_{>}(X,p)\tilde{\Pi}_{<}(X,p)]\,,
\end{align}
where we have introduced 
\begin{align}
\label{Ggtrless}
\tilde{G}_{\gtrless}(X,p)\equiv \tilde{G}_F(X,p)\pm
{\textstyle\frac12}\tilde{G}_\rho(X,p)\,
\end{align}
and their self--energy analogs $\tilde{\Pi}_\gtrless(X,p)$. 
The symmetry (antisymmetry) of the statistical (spectral) propagator with respect 
to permutation of its arguments and the definition  of the Wigner transformation 
imply that
\begin{align}
\label{GFGrhominp}
\tilde{G}_F(X,p)=\tilde{G}_F(X,-p),\,
\tilde{G}_\rho(X,p)=-\tilde{G}_\rho(X,-p)\,.
\end{align}
Therefore, for a single real scalar field, we have  
\begin{align}
\label{nofminp}
\tilde{G}_\gtrless(X,-p)=\tilde{G}_\lessgtr(X,p)\,,
\end{align}
and a similar relation for the self--energies.

\begin{figure}[!ht]
 \begin{center}
 \includegraphics[width=0.45\textwidth]{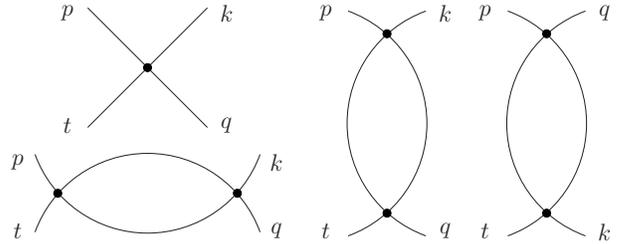}
\end{center}
\caption{\label{treeandoneloop} Feynman diagrams of 
$2\leftrightarrow 2$ scattering at tree and one--loop levels.}
\end{figure}

Explicit expressions for $\tilde{\Pi}_{\gtrless}(X,p)$ can be 
obtained after some algebra from Eqs.\,\eqref{PiFrho3loopWigner} 
and  \eqref{PiFrho4loopWigner}. For illustration purposes we   first derive $\Pi_{\gtrless}(x,y)$ and 
then perform the Wigner transformation. Using the decomposition 
\begin{align}
\hspace{-1.4mm}
G(x,y)=\theta(x^0-y^0)G_>(x,y)+\theta(y^0-x^0)G_<(x,y)
\end{align}
we obtain for the three--loop contribution 
\begin{align}
\Pi^{(3)}_\gtrless(x,y)=-\frac{\lambda^2}{6}G_\gtrless(x,y) 
G_\gtrless(x,y)G_\gtrless(x,y)\,.
\end{align}
Its Wigner transform reads 
\begin{align}
\label{Pi3gtrlessWigner}
\tilde{\Pi}^{(3)}_\gtrless(X,p)=&-\frac{\lambda^2}{6}\txtint d\Pi^4_kd\Pi^4_qd\Pi^4_t
(2\pi)^4 \delta_g(-p-t+k+q) \nonumber\\ 
&\times 
\tilde{G}_\lessgtr(X,t)\tilde{G}_\gtrless(X,k)\tilde{G}_\gtrless(X,q)
\end{align}
where we have used relation \eqref{nofminp}. It describes  
$2\leftrightarrow 2$ scattering and corresponds to the tree--level 
Feynman diagram in Fig.\,\ref{treeandoneloop}.

Expression for the four--loop contribution contains integration over the contour 
\begin{align}
\label{Pi4gtrless}
&\tilde{\Pi}^{(4)}_\gtrless(x,y)=\frac{\lambda^3}{2}G_\gtrless(x,y)\txtint \sqrt{-g}_z d^4z\,
\theta(z^0)\times\\
&[G_F(x,z)G_R(x,z)G^2_\gtrless(z,y)+G^2_\gtrless(x,z)G_A(z,y)G_F(z,y)].
\nonumber
\end{align}
After some algebra we obtain for the Wigner transform of \eqref{Pi4gtrless}
in the Boltzmann approximation (that is, with the Poisson brackets neglected) 
\begin{align}
\label{Pi4gtrlessWigner}
\tilde{\Pi}^{(4)}_\gtrless(&X,p)=\frac{\lambda^3}{2}\txtint d\Pi^4_kd\Pi^4_qd\Pi^4_t
(2\pi)^4 \delta_g(-p-t+k+q) \nonumber\\ &\times 
\tilde{G}_\lessgtr(X,t)\tilde{G}_\gtrless(X,k)\tilde{G}_\gtrless(X,q)
\,L(X,k+q)\,,
\end{align}
where
\begin{align}
\label{loopintegral}
L(X,p)\equiv\txtint d&\Pi^4_k  d\Pi^4_q (2\pi)^4 \delta_g(-p+k+q)\nonumber\\
&\times2\,\tilde{G}_F(X,k)\tilde{G}_h(X,q)\,.
\end{align}
From  \eqref{Pi4gtrlessWigner} it follows that  $L(X,p)$ is the 
same for the forward and inverse processes. As is demonstrated in Appendix \ref{oneloopclassic} 
it corresponds to the integrals of the one--loop Feynman 
diagrams in Fig.\,\ref{treeandoneloop}.

\begin{figure}[!ht]
 \begin{center}
 \includegraphics[width=0.45\textwidth]{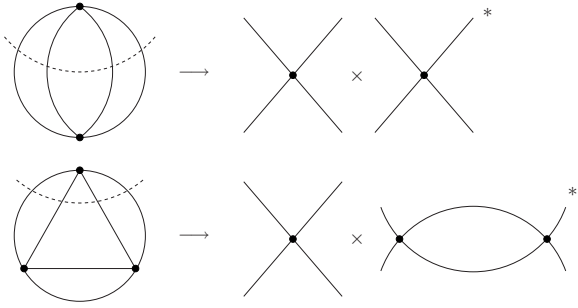}
\end{center}
\caption{\label{2PIcut} The correspondence between the diagrams contributing 
to the 2PI effective action and the contributions to the Boltzmann collision
terms.}
\end{figure}

Let us note here 
that the contribution(s) of a particular term of the 2PI effective 
action to the Boltzmann equation can be deduced by cutting the 2PI 
diagrams by a connected line in all possible ways. The three--loop
contribution, for instance, can be cut in only one way and the result 
can be represented as a product of two tree--level scattering diagrams.
The four--loop contribution can be cut in three equivalent ways and the result can be represented 
as a product of  tree--level and one--loop scattering diagrams, see 
Fig.\,\ref{2PIcut}.
There are two five--loop loop contributions to the effective action 
\cite{Berges:2004yj}. Applying the same procedure to one of them we would
obtain interference terms of two one--loop scattering diagrams and interference
of tree--level and two--loop scattering diagrams. Cutting the second, 
``eye'', diagram we would obtain interference of tree--level and two--loop
scattering diagrams and also interference of two $2\rightarrow 4$ 
diagrams.

The quasiparticle approximation \eqref{QPapprox} 
for $\tilde{G}_F(X,k)$ in \eqref{loopintegral} forces one of the intermediate 
states in the loop to be on the mass shell. On the contrary $\tilde{G}_h$, 
which describes the second intermediate state in the loop, vanishes on the 
mass shell. That is, the \textit{real intermediate state} contributions 
($2\rightarrow 2$ scattering into two \textit{on--shell} states followed 
by another $2\rightarrow 2$ scattering) are \textit{automatically} subtracted 
from the four--loop self--energies. 

Also note, that initial and final
states and on--shell intermediate states can be clearly 
distinguished in this formalism: the former ones are described by $\tilde{G}_\gtrless$ 
components, whereas the latter ones by $\tilde{G}_F$ or $\tilde{G}_\rho$ components.
Performing the integration 
and taking into account that one of the intermediate states is 
on--shell, we obtain the following expression for the loop integral:
\begin{align}
\label{oneloopexplicit}
L(X,p)&=\lim\limits_{\epsilon\rightarrow 0}\txtint 
\frac{d{\bf k}}{(2\pi)^3} \frac{2n(X,{\bf k})+1}{2E_k} \nonumber\\
\times&\left[
\frac{p^2-2pk}{(p^2-2pk)^2+\epsilon^2}+
\frac{p^2+2pk}{(p^2+2pk)^2+\epsilon^2}
\right]\,,
\end{align}
where $k=(E_k,{\bf k})$ is the on--shell four--momentum 
expressed in terms of the ``physical'' components:
$E_k\equiv k_0/\sqrt{g_{00}}$, etc.
In \eqref{oneloopexplicit} the background plasma 
``affects'' only one of the internal lines; the other one is off--shell 
and we can not associate the particle number density with it.

Next, we integrate the left-- and right--hand side of 
\eqref{BoltzmannPreliminary} over $p_0$ and choose the positive energy 
solution of \eqref{QPapprox} on the left--hand side. On the right--hand 
side both, the positive and the negative energy, solutions contribute. 
For positive $p_0$ momentum--energy conservation allows the following 
three combinations: 
\begin{align*}
a)\quad &k_0>0,\quad q_0>0,\quad t_0>0\,,\\
b)\quad &k_0>0,\quad q_0<0,\quad t_0<0\,,\\
c)\quad &k_0<0,\quad q_0>0,\quad t_0<0\,.
\end{align*}
As far as the three--loop self--energy \eqref{Pi3gtrlessWigner} is 
concerned, each combination leads to the same result, i.e.  an 
overall factor of $3$ appears. For the four--loop self--energy
the arising terms are not equal due to the presence of the loop integral $L$ in \eqref{Pi4gtrlessWigner}. 
Taking this into account and comparing 
\eqref{Pi3gtrlessWigner} and \eqref{Pi4gtrlessWigner} we see that 
in the 2PI formalism the effective coupling at nonzero particle number 
density at one--loop level contains a sum of three $L(X,p)$ functions 
with the arguments corresponding to $s$--, $t$-- and $u$--channel scattering:
\begin{align}
\label{Lambda}
\varLambda^2(X,k,q,t)&\equiv 
\lambda^2(1-\lambda [L(X,k+q)\nonumber\\
&+L(X,k-t)+L(X,q-t)])\,.
\end{align}

% However, we should note that a self--consistently dressed description 
% of the vertex in  $\lambda\varphi^4$ theory requires the use of the 4PI 
% effective action, which is beyond the scope of this paper.

After some algebra, the use of \eqref{nofminp} and redefinition of the momenta  
we  finally arrive at the Boltzmann equation for the distribution function:
\begin{widetext}
\begin{align}
\label{eqforn}
p^\alpha {\cal D}_\alpha  n(X,{\bf p})=&-\frac{\pi}{16}\int \frac{d{\bf k}}{(2\pi)^3}
\frac{d{\bf q}}{(2\pi)^3} \frac{d{\bf t} }{E_kE_qE_t}
\,\delta(E_p+E_t-E_q-E_k)\delta({\bf p}+{\bf t}-{\bf q}-{\bf k}\,)
\,\varLambda^2(X,{\bf k},{\bf q},{\bf t})
\nonumber\\
&
\times\left\{n(X,{\bf p})n(X,{\bf t})[n(X,{\bf k})+1][n(X,{\bf q})+1]-
[n(X,{\bf p})+1][n(X,{\bf t})+1]n(X,{\bf k})n(X,{\bf q})\right\}\,.
\end{align}
\end{widetext}

It is interesting, that the only remnant of the curved structure of  
space--time is the covariant derivative on the left--hand  side of the 
Boltzmann equation. In the case of greatest practical interest --  the 
Friedmann--Robertson--Walker universe -- it takes the  form 
\begin{align}
p^\alpha {\cal D}_\alpha n=\frac{E}{a}\left(\frac{\partial }{\partial\eta}-
\frac{{\bf p}^2}{E}{\cal H}\frac{\partial }{\partial E}\right)n\,,\quad
{\cal H}\equiv \frac{a'}{a}\,,
\end{align}
where $\eta$ is the conformal time. An integral form of the Boltzmann 
equation in the FRW universe as well as in a space--time with linearly 
perturbed FRW metric can be found, for instance, in \cite{Kartavtsev:2008fp}.

On the right--hand side, all the $\sqrt{-g}_X$ factors have disappeared 
due to  the introduction of the ``physical'' momenta and energies. 
In other words, the transition amplitudes in the scattering terms are independent 
of the space--time metric, which justifies  many earlier calculations. 
It is also remarkable that if only pointlike interactions (i.e. only the 
three--loop contribution to the 2PI effective action in the considered
case) are taken into account, Eq.\,\eqref{eqforn} coincides with the 
classical Boltzmann equation with the collision term calculated in 
\textit{vacuum}. The inclusion of four-- (and higher--loop) corrections 
to the effective potential induces further terms in the Boltzmann equation. 
These terms  correspond to the remnant space--time integrals in the 
self--energy and involve additional momentum integrals over the distribution 
functions.

\section{\label{summary}Summary and conclusions}

In this paper we have considered the dynamics of an 
out--of--equilibrium quantum system in a background 
gravitational field in the Schwinger--Keldysh formalism
. As one would 
expect, the resulting equations turned out to be covariant 
generalizations of their Minkowski--space counterparts. 

Remarkably, in the Boltzmann approximation 
the only remnant of the curved structure 
of the space--time is the covariant derivative on its left--hand 
side. The matrix elements of the scattering terms on the right--hand
side are independent of the metric. This justifies  
earlier calculations where this has been assumed implicitly.
Furthermore, if only the tree--level processes are 
taken into account, then the resulting equation coincides with 
the Boltzmann equation with the collision term calculated 
in \textit{vacuum}. Processes described by loop diagrams,
which induce corrections to the self--coupling, 
involve additional momentum integrals over the distribution 
functions, so that the resulting contributions differ from those 
calculated in vacuum. 

Interestingly, loop corrections, i.e. processes with intermediate 
\textit{off--shell} states, can be taken into account even if the 
quasiparticle Ansatz is applied. 
 As far as \textit{on--shell} intermediate 
states are concerned, there is a clear distinction between them 
and the initial and final states: the former ones are described by 
$\tilde{G}_F$ (or $\tilde{G}_\rho$) components, whereas the latter ones are given by 
$\tilde{G}_\gtrless$ components, see Eqs.\,\eqref{loopintegral} and \eqref{eqforn}. 
It is important that in the used formalism the problem of 
double--counting, which is cured
by a \textit{real intermediate state} subtraction procedure in the standard approach, 
does not arise at all.

For leptogenesis, this implies that whereas the 
washout processes described by contact interactions (they are 
present for instance in the supersymmetric extensions of the Standard Model) 
can be treated essentially classically, the correct treatment 
of the decay processes (which generate the asymmetry) and the scattering
processes mediated by the right--handed neutrino (which washout the 
asymmetry) requires the use of the Kadanoff--Baym approach.

Since the peculiarities of the calculation, related to the presence 
of a background gravitational field, are determined only by 
transformation properties of the fields -- scalar fields in the present 
case -- the developed formalism can be applied to arbitrary systems 
of scalar fields without any modifications. In \cite{Garny:2009rv} 
we study further implications of this formalism for leptogenesis 
and calculate the vertex contribution to the CP--violating parameter 
at nonzero particle densities in the framework of a toy model that 
qualitatively reproduces the features of popular leptogenesis models.
The analysis of the self--energy contributions to the CP--violating 
parameter will be performed in \cite{Hohenegger:2009b}.

\subsection*{Acknowledgements}
AH was supported by the ``Sonderforschungsbereich'' TR27. We thank 
Markus Michael M\"uller and Mathias Garny for sharing their insights in 
nonequilibrium quantum field theory and for very helpful discussions.

\begin{appendix}

\section{\label{oneloopclassic}$2\leftrightarrow 2$ scattering}
The tree--level amplitude of $2\leftrightarrow 2$ scattering (see 
Fig.\,\ref{treeandoneloop}) in Minkowski space--time  is given by 
\begin{align}
M^{tree}_{fi}=-i\lambda\,.
\end{align}
There are also three one--loop diagrams which contribute
to the scattering amplitude. Their contribution reads
\begin{align}
M^{loop}_{fi}=\frac{-\lambda^2}{2(2\pi)^4}\int 
\frac{d^4\xi\,d^4\eta\,\delta(-\sigma+\xi+\eta)}{
[\xi^2-M^2+i\epsilon][\eta^2-M^2+i\epsilon]}\,,
\end{align}
where  $\sigma$ is equal to $k+q$, to $k-t$ or $q-t$ 
(see Fig.\,\ref{treeandoneloop}). Because of the presence of 
the $\delta$--function one of the integrations (for instance,
over $\eta$) can be performed trivially. Calculating residues
of the integrand we can perform the integration over $d\xi_0$.
The result of the integration reads 
\begin{align}
\label{oneloopamplitude}
M^{loop}_{fi}=\frac{i\lambda^2}{2(2\pi)^3}\int 
\frac{d\xi^3}{2E_\xi}
\left[
\frac{1}{\xi^2+2\xi\sigma}+\frac{1}{\xi^2-2\xi\sigma}
\right]\,.
\end{align}
The quantity which enters the right--hand side of the 
Boltzmann equation is the amplitude modulo squared.
To leading order in small $\lambda$ it is given by 
\begin{align}
\label{sumoftheamplitudes}
|M_{fi}|^2=\lambda^2(1-&\lambda[L^{vac}(k+q)\nonumber\\
&+L^{vac}(k-t)+L^{vac}(q-t)])\,,
\end{align}
where $L^{vac}(\sigma)$ coincides with \eqref{oneloopexplicit}
if $n(X,{\bf k})$ and $\epsilon$ are set to zero. 
The former condition arises from the fact that in this Appendix we
calculate the scattering amplitudes in \textit{vacuum}, whereas 
the latter one is related to the fact that we have not subtracted 
the contributions of  \textit{real intermediate states} to the 
one--loop amplitude. Comparing \eqref{sumoftheamplitudes} with 
\eqref{Lambda} we conclude that $L(X,p)$ indeed describes the integrals 
of the one--loop diagrams.
\end{appendix}

%=============================================================================
%\bibliographystyle{apsrev}
%\bibliography{kb}
% =============================================================================

\end{document}